\documentclass[a4paper]{article}
\pdfoutput=1 
\usepackage{subcaption}
\usepackage{INTERSPEECH2020}
\usepackage{makecell}
\usepackage[skip=3pt]{caption} 
\usepackage{multirow}
\usepackage{color}
\usepackage{amsmath}
\usepackage{cite}
\usepackage{url}

\usepackage{breakurl}
\usepackage{algorithmic}
\usepackage{algorithm}

\title{Unifying Cosine and PLDA Back-ends for Speaker Verification}
\name{Zhiyuan Peng$^{1,2}$, Xuanji He$^2$, Ke Ding$^2$, Tan Lee$^1$, Guanglu Wan$^2$}
\address{
  $^1$Department of Electronic Engineering, The Chinese University of Hong Kong\\
  $^2$Meituan}
 \email{jerrypeng1937@gmail.com, \{hexuanji,dingke02,wanguanglu\}@meituan.com, tanlee@ee.cuhk.edu.hk}

\begin{document}

\maketitle

\begin{abstract}

State-of-art speaker verification (SV) systems use a back-end model to score the similarity of speaker embeddings extracted from a neural network model. The commonly used back-end models are the cosine scoring and the probabilistic linear discriminant analysis (PLDA) scoring. With the recently developed neural embeddings, the theoretically more appealing PLDA approach is found to have no advantage against or even be inferior the simple cosine scoring in terms of SV system performance. This paper presents an investigation on the relation between the two scoring approaches, aiming to explain the above counter-intuitive observation. It is shown that the cosine scoring is essentially a special case of PLDA scoring. In other words, by properly setting the parameters of PLDA, the two back-ends become equivalent. As a consequence, the cosine scoring not only inherits the basic assumptions for the PLDA but also introduces additional assumptions on the properties of input embeddings. Experiments show that the dimensional independence assumption required by the cosine scoring contributes most to the performance gap between the two methods under the domain-matched condition. When there is severe domain mismatch and the dimensional independence assumption does not hold, the PLDA would perform better than the cosine for domain adaptation.


\end{abstract}

\section{Introduction}\label{sec:intro} 

Speaker verification (SV) is the task of verifying the identity of a person from the characteristics of his or her voice. It has been widely studied for decades with significant performance advancement.  State-of-the-art SV systems are predominantly embedding based, comprising a front-end embedding extractor and a back-end scoring model. The front-end module transforms input speech into a compact embedding representation of speaker-related acoustic characteristics. The back-end model computes 
the similarity of two input speaker embeddings and determines whether they are from the same person.

There are two commonly used back-end scoring methods. One is the cosine scoring, which assumes the input embeddings are angularly discriminative. The SV score is defined as the cosine similarity of two embeddings $x_1$ and $x_2$, which are mean-subtracted and length-normalized \cite{garcia2011analysis}, i.e.,
    \begin{align}
        & x_i \leftarrow \frac{x_i - \mu}{||x_i - \mu||_2}, \text{ for } i = 1, 2 \label{eq:pre_norm} \\
        & S_{\text{cos}}(x_1, x_2) = x_1^T x_2 \label{eq:cos_score}
    \end{align}
The other method of back-end scoring is based on probabilistic linear discriminant analysis (PLDA) \cite{ioffe2006probabilistic}. It takes the assumption that the embeddings (also mean-subtracted and length-normalized) are in general Gaussian distributed. 

It has been noted that the standard PLDA back-end performs significantly better than the cosine back-end on conventional i-vector embeddings \cite{dehak2010front}. Unfortunately, with the powerful neural speaker embeddings that are widely used nowadays \cite{zeinali2019but}, the superiority of PLDA vanishes and even turns into inferiority. This phenomenon has been evident in our experimental studies, especially when the front-end is trained with the additive angular margin softmax loss\cite{deng2019arcface,xiang2019margin}.

The observation of PLDA being not as good as the cosine similarity is against the common sense of the back-end model design. Compared to the cosine, PLDA has more learnable parameters and incorporates additional speaker labels for training.
Consequently, PLDA is generally considered to be more effective in discriminating speaker representations. 
This contradiction between experimental observations and theoretical expectation deserves thoughtful investigations on PLDA. In \cite{li2019gaussian,zhang2019vae,cai2020deep}, Cai et al  argued that the problem should have arise from the neural speaker embeddings. It is noted that embeddings extracted from neural networks tend to be non-Gaussian for individual speakers and the distributions across different speakers are non-homogeneous. These irregular distributions cause  the performance degradation of verification systems with the PLDA back-end. In relation to this perspective, a series of regularization approaches have been proposed to force the neural embeddings to be homogeneously Gaussian distributed, e.g., Gaussian-constrained loss \cite{li2019gaussian}, variational auto-encoder\cite{zhang2019vae} and discriminative normalization flow\cite{cai2020deep,li2020neural}.

In this paper, we try to present and substantiate a very different point of view from that in previous research. We argue that the suspected irregular distribution of speaker embeddings does not necessarily contribute to the inferiority of PLDA versus the cosine.
Our view is based on the evidence that the cosine can be regarded as a special case of PLDA. This is indeed true but we have not yet found any work mentioning it. Existing studies have been treating the PLDA and the cosine scoring methods separately. We provide a short proof to unify them. It is noted that the cosine scoring, as a special case of PLDA, also assumes speaker embeddings to be homogeneous Gaussian distributed. Therefore, if the neural speaker embeddings are distributed irregularly as previously hypothesized, both back-ends should exhibit performance degradation.

By unifying the cosine and the PLDA back-ends, it can be shown that the cosine scoring puts stricter assumptions on the embeddings than PLDA. Details of these assumptions are explained in Section \ref{sec:cosine_is_plda}. Among them, the dimensional independence assumption is found to play a key role in explaining the performance gap between the two back-ends. It is evidenced by incorporating the dimensional independence assumption into the training of PLDA, leading to the diagonal PLDA (DPLDA). This variation of PLDA shows a significant performance improvement under the domain-matched condition. 
However, when severe domain mismatch exists and back-end adaptation is needed, PLDA performs better than both the cosine and DPLDA. This is because the dimension independence assumption does not hold. Analysis on the between-/within-class covariance of speaker embeddings supports these statements.

\section{Review of PLDA}
Theoretically PLDA is a probabilistic extension to the classical linear discriminant analysis (LDA)\cite{balakrishnama1998linear}. It incorporates a Gaussian prior on the class centroids in LDA. Among the variants of PLDA, the two-covariance PLDA\cite{sizov2014unifying} has been commonly used in speaker verification systems. A straightforward way to explain two-covariance PLDA is by using probabilistic graphical model\cite{jordan2003introduction}.

\subsection{Modeling}

\begin{figure}[htbp]
    \centering
    \includegraphics[width=0.3\textwidth]{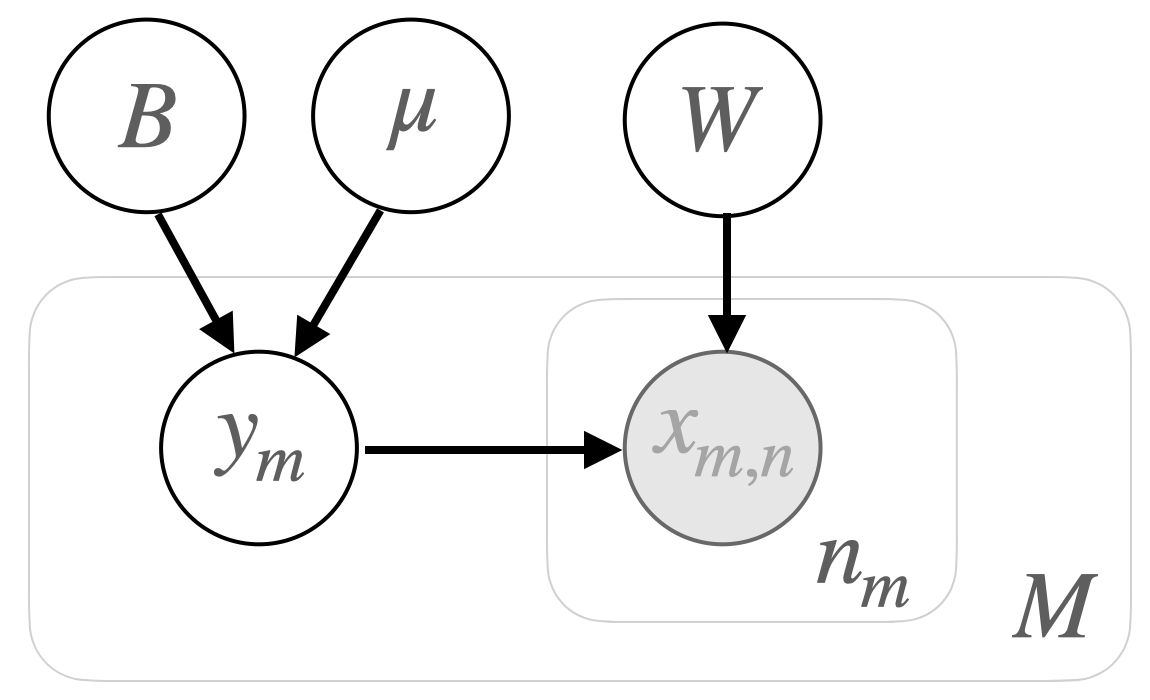}
    \caption{The probabilistic graphical model of two-covariance PLDA}
    \label{fig:plda_pgm}
\end{figure}

Consider $N$ speech utterances coming from $M$ speakers, where the $m$-th speaker is associated with $n_m$ utterances.
With a front-end embedding extractor, each utterance can be represented by an embedding of $D$ dimensions. The embedding of the $n$-th utterance from the $m$-th speaker is denoted as $x_{m,n}$. Let $\mathcal{X} =\{x_{m,n}\}_{1,1}^{M, n_m}$ represent these per-utterance embeddings.
Additionally, PLDA supposes the existence of per-speaker embeddings $\mathcal{Y} = \{y_m\}_{m=1}^{M}$. They are referred to as latent \textit{speaker identity variables} in \cite{brummer2010speaker}.

With the graphical model shown in Fig.\ref{fig:plda_pgm}, these embeddings are generated as follows,
\begin{itemize}
    \item Randomly draw the per-speaker embedding $y_m \sim \mathcal{N} (y_m; \mu, B^{-1})$, for $m = 1,\cdots,M$;
    \item Randomly draw the per-utterance embedding $x_{m,n} \sim \mathcal{N} (x_{m,n}; y_m, W^{-1})$, for $n = 1, \cdots, n_m$.
\end{itemize}
where $\theta = \{\mu, B, W\}$ denotes the model parameters of PLDA. Note that $B$ and $W$ are precision matrices.
The joint distribution $p_\theta(\mathcal{X}, \mathcal{Y})$ can be derived as,

\begin{equation}
  \begin{split}
    p_\theta(\mathcal{X}, \mathcal{Y})  \propto \text{exp}(-\frac{1}{2}\sum_{m=1}^M &\Bigl[(y_m -\mu)^T B (y_m-\mu) \\  
    +&\sum_{n=1}^{n_m} (x_{m,n}-y_m)^T W (x_{m,n}-y_m)\Bigl] )
\end{split}
\end{equation}

\subsection{Training}
Estimation of PLDA model parameters can be done with the iterative E-M algorithm, as described in Algorithm \ref{alg:plda_em}. The algorithm requires initialization of model parameters. In kaldi\cite{povey2011kaldi}, the initialization strategy is to set $B=W=I$ and $\mu=0$.

\begin{algorithm}[H]
\caption{E-M training of two-covariance PLDA}\label{alg:plda_em}
\begin{algorithmic}
 \STATE \textbf{Input}: per-utterance embeddings $\mathcal{X} =\{x_{m,n}\}_{1,1}^{M, n_m}$
 \STATE \textbf{Initialization}: $B=W=I, \mu = 0$
 \REPEAT
 \STATE \textbf{(E-step):} Infer the latent variable $y_m | \mathcal{X}$
 \STATE \hspace{0.3cm} $L_m = B + n_m W$
 \STATE \hspace{0.3cm} $y_m |\mathcal{X} \sim \mathcal{N} (L_m^{-1} (B\mu + W \sum_{n=1}^{n_m}x_{m,n}), L_m^{-1})$
 
 \STATE \textbf{(M-step):} Update $\theta$ by $\max_\theta \mathbb{E}_{\mathcal{Y}} \log p_\theta (\mathcal{X}, \mathcal{Y})$
 \STATE \hspace{0.3cm} $\mu = \frac{1}{M} \sum_m \mathbb{E}[y_m|\mathcal{X}]$
 \STATE \hspace{0.3cm} $B^{-1} = \frac{1}{M} \sum_m \mathbb{E} [ y_m y_m^T| \mathcal{X} ] - \mu \mu^T$
 \STATE \hspace{0.3cm} $W^{-1} = \frac{1}{N} \sum_m \sum_n \mathbb{E} [ (y_m-x_{m,n})(y_m-x_{m,n})^T | \mathcal{X} ]$
 \UNTIL{Convergence}\;
 \STATE \textbf{Return} $B, W, \mu$

\end{algorithmic}
\label{alg1}
\end{algorithm}

\subsection{Scoring}
Assuming the embeddings are mean-subtracted and length-normalized, we let $\mu \approx 0$ to simplify the scoring function. Given two per-utterance embeddings $x_i, x_j$, the PLDA generates a log-likelihood ratio (LLR) that measures the relative likelihood of the two embeddings coming from the same speaker. The LLR is defined as,

\begin{align}
    S_{\text{PLDA}}(x_i, x_j) &= \log \frac{p(x_i, x_j | \mathcal{H}_1)}{p(x_i, x_j | \mathcal{H}_0)} \nonumber \\
    &= \log \frac{p(x_i, x_j)}{p(x_i)p(x_j)} \label{eq:llr}
\end{align}
where $\mathcal{H}_1$ and $\mathcal{H}_0$ represent the same-speaker and different-speaker hypotheses. To derive the score function, without loss of generality, consider a set of $n_1$ embeddings $\mathcal{X}_1 = \{x_{1,n}\}_{n=1}^{n_1}$ that come from the same speaker. It can be proved that 

\begin{align}\label{eq:lemma_px}
    &\log p(\mathcal{X}_1) = \\
    &\frac{1}{2}\left( n_1^2 \mu_1^T W(B+n_1W)^{-1}W\mu_1 - \sum_{n=1}^{n_1} x_{1,n}^T W x_{1,n} \right. \nonumber\\
    &+\log |B| + n_1\log |W| - \log |B + n_1 W| - n_1 D \log(2\pi)\biggr) \nonumber
\end{align}
where $\mu_1 = \frac{1}{n_1} \sum_{n=1}^{n_1} x_{1,n}$. By applying Eq.\ref{eq:lemma_px} into Eq.\ref{eq:llr}, the LLR can be expressed as
\begin{equation}\label{eq:plda_score}
    S_{\text{PLDA}}(x_i, x_j) \dot{=} \frac{1}{2}\left(x_i^TQx_i + x_j^TQx_j + 2x_i^TPx_j\right)
\end{equation}
where $\dot{=}$ means equivalence up to a negligible additive constant, and
\begin{align}
    Q &= W((B+2W)^{-1} - (B+W)^{-1})W \\
    P &= W (B+2W)^{-1}W
\end{align}
Note that $Q \prec 0$ and $P + Q \succeq 0$.

\section{Cosine as a typical PLDA}\label{sec:cosine_is_plda}

Relating Eq.\ref{eq:plda_score} to Eq.\ref{eq:cos_score} for the cosine similarity measure, it is noted that when $-Q = P = I$, the LLR of PLDA degrades into the cosine similarity, as $x_i^Tx_i = 1$. It is also noted that the condition of $-Q = P = I$ is not required. PLDA is equivalent to the cosine if and only if $Q = \alpha I$ and $P = \beta I$, where $\alpha < 0, \alpha +\beta \geq 0$.

Given $W \succ 0$, we have
\begin{align}
    W &= \frac{\beta(\beta-\alpha)}{-\alpha} I\\
    B &= \frac{\beta (\beta +\alpha)(\beta-\alpha)}{\alpha^2}I
\end{align}
Without loss of generality, we let $W = B = I$. In other words, the cosine is a typical PLDA with both within-class covariance $W^{-1}$ and between-class covariance $B^{-1}$ fixed as an identity matrix.

So far we consider only the simplest pairwise scoring. In the general case of many-vs-many scoring, the PLDA and cosine are also closely related. For example, let us consider two sets of embeddings $\mathcal{X}_1$ and $\mathcal{X}_2$ of size $K_1$ and $K_2$, respectively. Their centroids are denoted by $\mu_1$ and $\mu_2$. It can be shown,
\begin{align}
S_{\text{PLDA}}(\mathcal{X}_1, \mathcal{X}_2) &= \frac{K_1 K_2}{1+K_1 + K_2} S_{\text{cos}}(\mu_1, \mu_2) + \frac{1}{2}C(K_1, K_2)  \\
C(K_1, K_2) &= \frac{K_1^2 + K_2^2}{1+K_1 +K_2} - \frac{K_1^2}{1+K_1} -\frac{K_2^2}{1+K_2} \nonumber \\
 & + \log(1+\frac{K_1K_2}{1+K_1 + K_2})
\end{align}
under the condition of $W = B = I$. The term $C(K_1, K_2)$ depends only on $K_1$ and $K_2$.

This has shown that the cosine puts more stringent assumptions than PLDA on the input embeddings. These assumptions are:
\begin{enumerate}
    \item (\textbf{dim-indep}) Dimensions of speaker embeddings are mutually uncorrelated or independent;
    \item Based on 1), all dimensions share the same variance value.
\end{enumerate}
As the embeddings are assumed to be Gaussian, dimensional uncorrelatedness is equivalent to dimensional independence.

\subsection{Diagonal PLDA}\label{sec:dplda}
With Gaussian distributed embeddings, the \textit{dim-indep} assumption implies that speaker embeddings have diagonal covariance. To analyse the significance of this assumption to the performance of SV backend, a diagonal constraint is applied to updating $B$ and $W$ in Algorithm \ref{alg:plda_em}, i.e.,
\begin{align}
    B^{-1} &= \text{diag}(\frac{1}{M} \sum_m \mathbb{E} [ y_m^{\circ 2}| \mathcal{X} ] - \mu^{\circ 2}) \label{eq:dplda1}\\
    W^{-1} &= \text{diag}(\frac{1}{N} \sum_m \sum_n \mathbb{E} [ (y_m-x_{m,n})^{\circ 2} | \mathcal{X} ])\label{eq:dplda2}
\end{align}
where $\circ 2$ denotes the Hadamard square. The PLDA trained in this way is named as the diagonal PLDA (DPLDA). The relationship between DPLDA and PLDA is similar to that between the diagonal GMM and the full-covariance GMM.

\section{Experimental setup}
Experiments are carried out with the Voxceleb1+2 \cite{nagrani2017voxceleb} and the CNCeleb1 databases\cite{li2022cn}. A vanilla ResNet34\cite{chung2020defence} model is trained with 1029K utterances from 5994 speakers in the training set of Voxceleb2. Following the state-of-the-art training configuration\footnote{https://github.com/TaoRuijie/ECAPA-TDNN}, data augmentation with speed perturbation, reverberation and spectrum augmentation\cite{park2019specaugment} is applied. The AAM-softmax loss\cite{deng2019arcface} is adopted to produce angular-discriminative speaker embeddings.

The input features to ResNet34 are 80-dimension filterbank coefficients with mean normalization over a sliding window of up to 3 seconds long. Voice activity detection is carried out with the default configuration in kaldi\footnote{https://github.com/kaldi-asr/kaldi/blob/master/egs/voxceleb/v2/conf}.
The front-end module is trained to generate 256-dimension speaker embeddings, which are subsequently mean-subtracted and length-normalized. The PLDA backend is implemented in kaldi and modified to the DPLDA according to Eq. \ref{eq:dplda1}-\ref{eq:dplda2}.

Performance evaluation is carried out on the test set in VoxCeleb1 and CNCeleb1. The evaluation metrics are equal error rate (EER) and decision cost function (DCF) with $p_\text{tar} = 0.01$ or $0.001$.

\subsection{Performance comparison between backends}
As shown in Table \ref{tab:cos_plda_voxceleb}, the performance gap between cosine and PLDA backends can be observed from the experiment on VoxCeleb. Cosine outperforms PLDA by relatively improvements of $51.61\%$ in terms of equal error rate (EER) and $50.73\%$ in terms of minimum Decision Cost Function with $P_{\text{tar}} = 0.01$ (DCF$0.01$). 
The performance difference becomes much more significant with DCF$0.001$, e.g., $0.3062$ by PLDA versus $0.1137$ by the cosine. Similar results are noted on other test sets of VoxCeleb1 ((not listed here for page limit)).

The conventional setting of using LDA to preprocess raw speaker embeddings before PLDA is evaluated. It is labelled as \textit{LDA+PLDA} in Table \ref{tab:cos_plda_voxceleb}. Using LDA appears to have a negative effect on PLDA. This may be due to the absence of the \textit{dim-indep} constraint on LDA. We argue that it is unnecessary to apply LDA to regularize the embeddings. The commonly used LDA preprocessing is removed in the following experiments. 

\begin{table}[htbp]
    \centering
    \caption{Comparison of backends on VoxCeleb. }
    \begin{tabular}{c||c|c|c}
    \toprule
        & EER\% & DCF0.01 & DCF0.001  \\
    \midrule
    \midrule
        cos & \textbf{1.06} & \textbf{0.1083} & \textbf{0.1137} \\
        PLDA & 1.86 & 0.2198 & 0.3062 \\
        LDA+PLDA & 2.17 & 0.2476 & 0.3715 \\
    \midrule
        DPLDA & 1.11 & 0.1200 & 0.1426 \\
    \bottomrule
    \end{tabular}
    \label{tab:cos_plda_voxceleb}
\end{table}

The DPLDA incorporates the \textit{dim-indep} constraint into PLDA training. 
As shown in Table \ref{tab:cos_plda_voxceleb}, it improves the EER of PLDA from $1.86\%$ to $1.11\%$, which is comparable to cosine. This clearly confirms the importance of \textit{dim-indep}.

\subsection{Performance degradation in Iterative PLDA training}
\begin{figure}[htbp]
    \centering
    \includegraphics[width=0.4\textwidth]{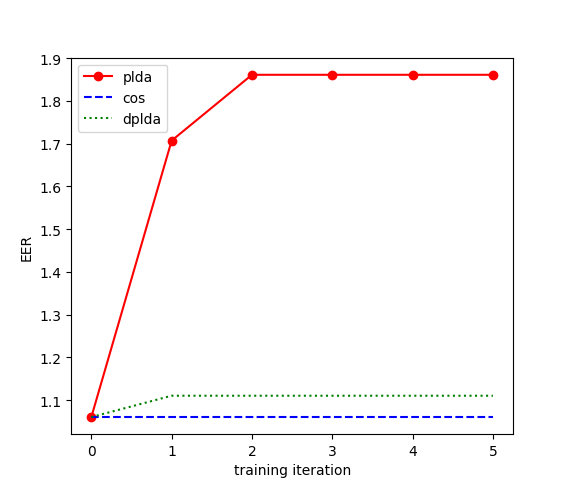}
    \caption{PLDA gets worse in its iterative EM training}
    \label{fig:plda_training_eer}
\end{figure}

According to the derivation in Section \ref{sec:cosine_is_plda}, PLDA implemented in Algorithm \ref{alg:plda_em} is initialized as the cosine, e.g., $B=W=I$. However, the PLDA has been shown to be inferior to the cosine by the results in Table \ref{tab:cos_plda_voxceleb}. Logically it would be expected that the performance of PLDA degrades in the iterative EM training. Fig \ref{fig:plda_training_eer} shows the plot of EERs versus number of training iterations.
Initially PLDA achieves exactly the same performance as cosine. In the first iteration, the EER seriously increases from 1.06\% to 1.707\%. For DPLDA, the \textit{dim-indep} constraint shows an effect of counteracting the degradation.
\vspace{-2mm}

\subsection{When domain mismatch exists}

The superiority of cosine over PLDA has been evidenced on the VoxCeleb dataset, of which both training and test data come from the same domain, e.g., interviews collected from YouTube. In many real-world scenarios, domain mismatch between training and test data commonly exists. A practical solution is to acquire certain amount of in-domain data and update the backend accordingly. The following experiment is to analyse the effect of domain mismatch on the performance of backend models. 

The CNCeleb1 dataset is adopted as the domain-mismatched data. It is a multi-genre dataset of Chinese speech with very different acoustic conditions from VoxCeleb. The ResNet34 trained on VoxCeleb is deployed to exact embeddings from the utterances in CNCeleb1. 
The backends are trained and evaluated on the training and test embeddings of CNCeleb1.

 As shown in Table\ref{tab:backend_cnceleb}, the performance of both cosine and DPLDA are inferior to PLDA. Due to that the \textit{dim-indep} assumption no longer holds, the diagonal constraint on covariance does not bring any performance improvement to cosine and DPLDA. 
\begin{table}[htbp]
    \centering
    \caption{Comparison of backends on CNCeleb1}
    \begin{tabular}{c||c|c|c}
    \toprule
        & EER\% & DCF0.01 & DCF0.001 \\
    \hline
    \hline
       cos & 10.11 & 0.5308 & 0.7175 \\
       PLDA & \textbf{8.90} & \textbf{0.4773} & \textbf{0.6331} \\ 
    \hline
        DPLDA & 10.24 & 0.5491 & 0.8277 \\
    \bottomrule
    \end{tabular}
    \label{tab:backend_cnceleb}
\end{table}

\subsection{Analysis of between-/within-class covariances }
To analyze the correlation of individual dimensions of the embeddings, the between-class and within-class covariances, $B_0^{-1}$ and $W_0^{-1}$, are computed as follows,
\begin{align}
    B_0^{-1} &= \frac{1}{M} \sum_M n_m y_m y_m^T - \mu_0\mu_0^T \\
    W_0^{-1} &= \frac{1}{M} \sum_{m=1}^M \sum_{n=1}^{n_m} (x_{m,n} - y_m)(x_{m,n} - y_m)^T
\end{align}
where $\mu_0 = \frac{1}{N}\sum_{m=1}^{M}\sum_{n=1}^{n_m} x_{m,n}$ and $y_m = \frac{1}{n_m}\sum_{n=1}^{n_m} x_{m,n}$. These are the training equations of LDA and closely related to the M-step of PLDA. Note that for visualization, the elements in $B_0^{-1}$ and $W_0^{-1}$ are converted into their absolute value. 

\begin{figure}[htbp]
    \centering
    \includegraphics[width=0.5\textwidth]{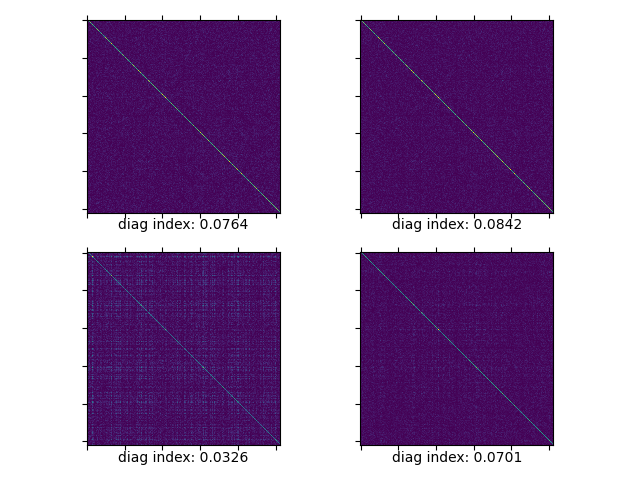}
    \caption{between-class (left) and within-class (right) covariance of embeddings on the training data of VoxCeleb (top) and CN-Celeb (bottom). The diagonal index is computed as $\text{trace}(G)/\text{sum}(G)$ for a non-negative covariance matrix $G$.
    \label{fig:btw_within_covar_show}}
\end{figure}

In Fig.\ref{fig:btw_within_covar_show}, both between-class and within-class covariances show clearly diagonal patterns, in the domain-matched case (plot on the top). This provides additional evidence to support the \textit{dim-indep} assumption aforementioned. However, this assumption would be broken with strong domain-mismatched data in CNCeleb. As shown by the two sub-plots in the bottom of Fig \ref{fig:btw_within_covar_show}, even though the within-class covariance plot on the right shows a nice diagonal pattern, it tends to vanish for the between-class covariance (plot on the left). 
Off-diagonal elements have large absolute value and the dimension correlation pattern appears, suggesting the broken of \textit{dim-indep}. The numerical measure of diagonal index also confirms this observation.

\vspace{-1mm}
\section{Conclusion}\label{conclusions}

The reason why PLDA appears to be inferior to the cosine scoring with neural speaker embeddings has been exposed with both theoretical and experimental evidence. It has been shown that the cosine scoring is essentially a special case of PLDA. Hence, the non-Gaussian distribution of speaker embeddings should not be held responsible for explaining the performance difference between the PLDA and cosine back-ends. Instead, it should be attributed to the dimensional independence assumption made by the cosine, as evidenced in our experimental results and analysis. Nevertheless, this assumption fits well only in the domain-matched condition. When severe domain mismatch exists, the assumption no longer holds and PLDA can work better than the cosine.
Further improvements on PLDA need to take this assumption into consideration.
It is worth noting that the AAM-softmax loss should have the benefit of regularizing embeddings to be homogeneous Gaussian, considering good performance of the cosine scoring.

\bibliographystyle{IEEEtran}

\bibliography{me}

\end{document}